\documentclass[]{spie}  

\usepackage{amsmath,amsfonts,amssymb}
\usepackage{graphicx}
\usepackage[colorlinks=true, allcolors=blue]{hyperref}

\title{The infrared imaging spectrograph (IRIS) for TMT: electronics-cable architecture}

\author[a]{A.C. Trapp}
\author[a]{James Larkin}
\author[a]{Ken Magnone}
\author[b]{Timothee Greffe}
\author[c]{Tim Hardy}
\author[c]{Jennifer Dunn}
\author[d]{Eric Chisholm}
\author[a]{Chris Johnson}
\author[e]{Ryuji Suzuki}

\affil[a]{UCLA Department of Physics and Astronomy, 430 Portola Plaza, Los Angeles, CA, USA}
\affil[b]{Caltech, 1216 E California Blvd, Pasadena, CA 91125}
\affil[c]{NRC Herzberg Astronomy and Astrophysics Research Centre}
\affil[d]{TMT, 100 West Walnut Street, Pasadena, CA, 91124}
\affil[e]{NAOJ, Osawa, Mitaka, Tokyo, Japan}

\begin{document} 
\maketitle

\begin{abstract}
The InfraRed Imaging Spectrograph (IRIS) is a first-light instrument for the Thirty Meter Telescope (TMT). It combines a diffraction limited imager and an integral field spectrograph. This paper focuses on the electrical system of IRIS. With an instrument of the size and complexity of IRIS we face several electrical challenges. Many of the major controllers must be located directly on the cryostat to reduce cable lengths, and others require multiple bulkheads and must pass through a large cable wrap. Cooling and vibration due to the rotation of the instrument are also major challenges. We will present our selection of cables and connectors for both room temperature and cryogenic environments, packaging in the various cabinets and enclosures, and techniques for complex bulkheads including for large detectors at the cryostat wall.  
\end{abstract}


\section{INTRODUCTION}
\label{sec:intro}  
The InfraRed Imaging Spectrograph (IRIS\cite{Larkin18}) is a first-generation instrument for the Thirty Meter Telescope (TMT\cite{Sanders13}). A combination of the On-Instrument Wave Front Sensor (OIWFS\cite{Dunn14}) and the TMT adaptive optics system NFIRAOS\cite{Herriot14} will allow IRIS to reach the diffraction limit of TMT at wavelengths longer than 1 micron. IRIS combines an imager and an integral field spectrograph operating between 0.8 to 2.5 microns. The imager is composed of four 4K by 4K Teledyne detectors (Hawaii 4RG) with 4 mas pixels and a combined 34 x 34 $\text{arcsec}^2$ field of view. The integral field spectrograph takes $\sim10,000$ spectra simultaneously with spectral resolution $R\sim4000\ \text{to}\ R\sim8000$ on four spatial scales from 4 mas to 50 mas with fields of view of 0.46 x 0.51 $\text{arcsec}^{2}$ to 2.2 x 4.6 $\text{arcsec}^{2}$ respectively. The science applications of IRIS span from our own Solar System to the most distant galaxies in the Universe.

    \begin{figure} [ht]
    \begin{center}
    \begin{tabular}{c} 
    \includegraphics[width=15cm]{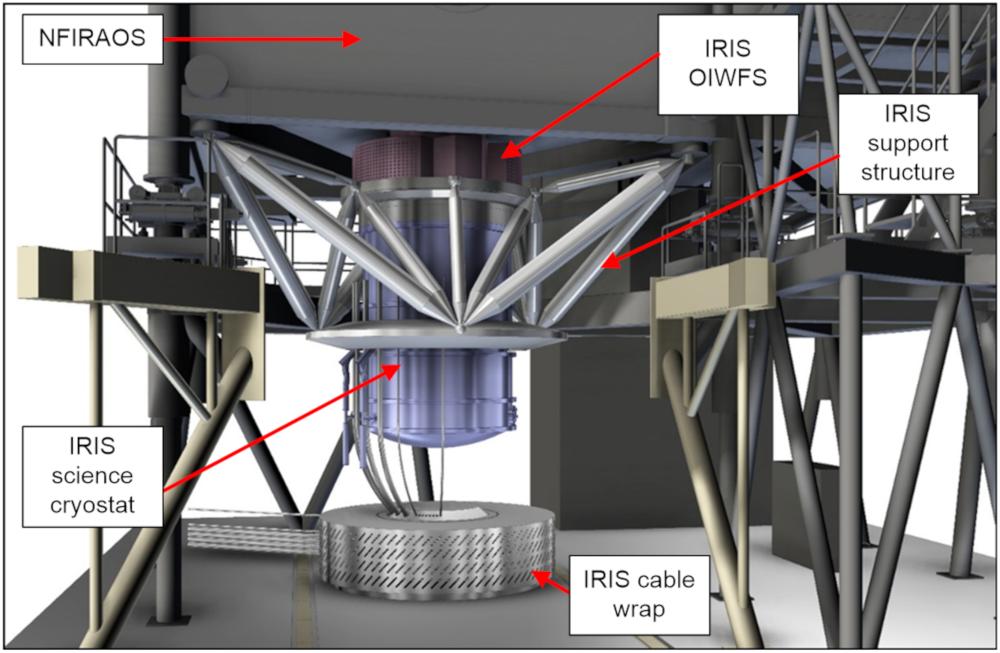}
    \end{tabular}
    \end{center}
    \caption[cryostat] 
    { \label{fig:cryostat}
IRIS is mounted to the underbelly of TMT’s AO system NFIRAOS by the IRIS support structure. After exiting NFIRAOS, light passes through the On Instrument Wave Front Sensor (OIWFS), which allows for tip-tilt, focus, and plate scale corrections, and finally through a window into the science cryostat, which has both an imaging and integral field spectrometer mode. The cable wrap sits below the cryostat on the nasmyth platform, and connects to the nasmyth cabinets (not shown here). The current design also includes a bulkhead around the bottom rim of the cryostat (not shown here) which acts as a cable break for easier installation of the instrument. Also not shown here are the detector control boxes (see Fig.~\ref{fig:leach}) which will be mounted on the sides of the cryostat.}
    \end{figure}
    
    \begin{figure} [hb]
    \begin{center}
    \begin{tabular}{c} 
    \includegraphics[width=6cm]{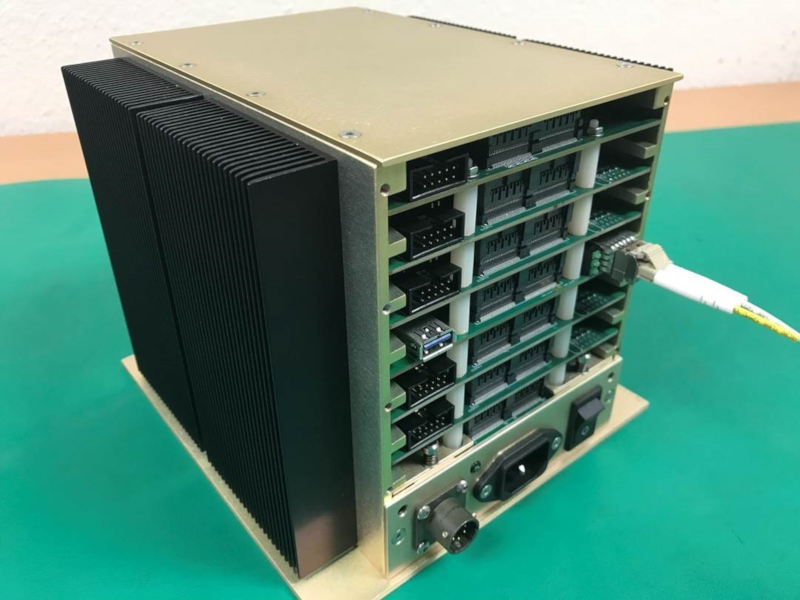}
    \end{tabular}
    \end{center}
    \caption[leach] 
    { \label{fig:leach} 
An example of an ARC box; similar enclosures will be attached to the side of the science cryostat and will house detector control electronics.}
    \end{figure}
    
    \begin{figure} [hb]
    \begin{center}
    \begin{tabular}{c} 
    \includegraphics[width=15cm]{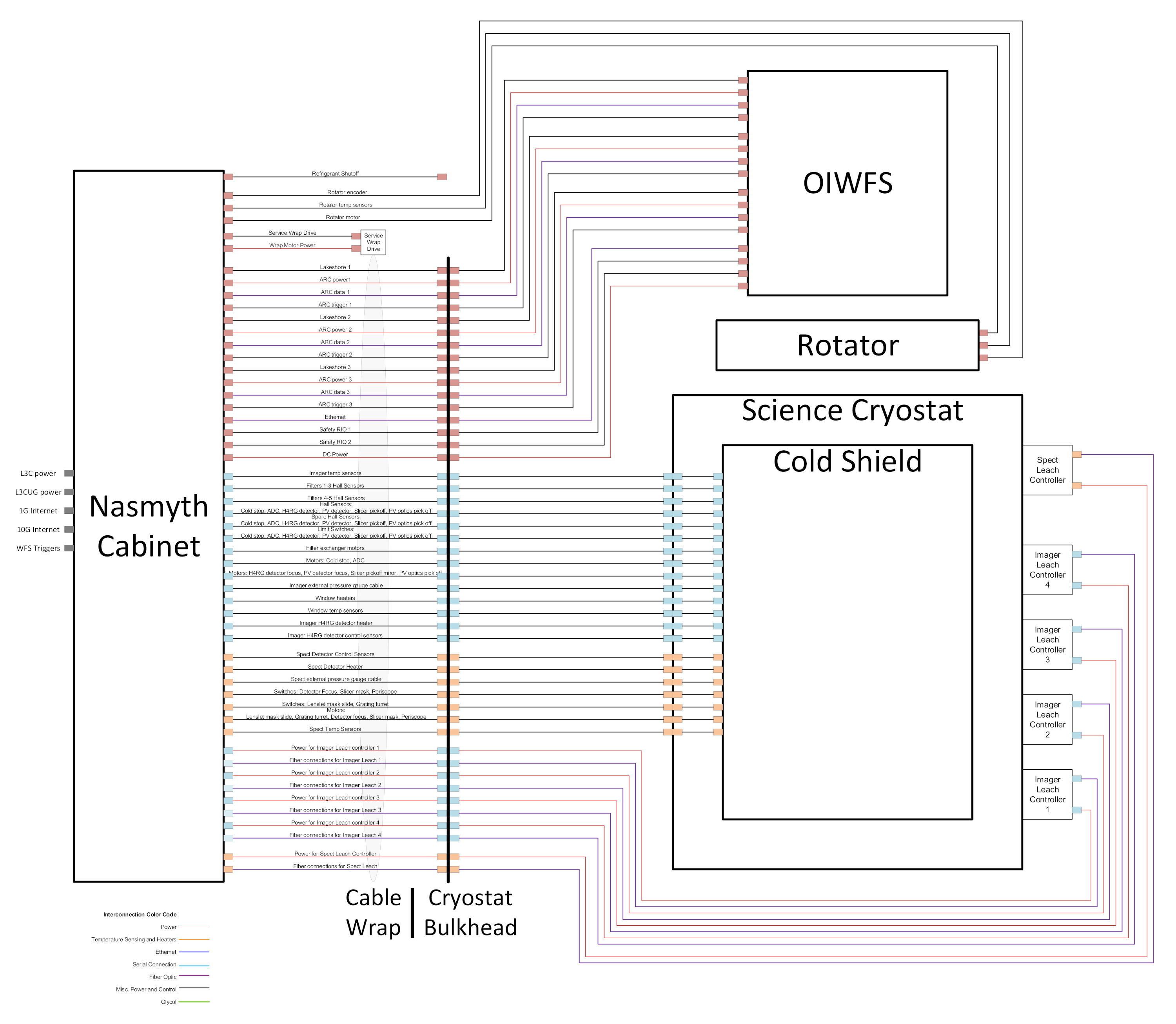}
    \end{tabular}
    \end{center}
    \caption[over] 
    { \label{fig:over} 
A high-level schematic diagram of Fig.~\ref{fig:cryostat} showing the large number of cable connections required to control IRIS, as well as their paths from the nasmyth cabinets to the cryostat, rotator, or OIWFS.}
    \end{figure}
    
    \begin{figure} [ht]
    \begin{center}
    \begin{tabular}{c} 
    \includegraphics[width=15cm]{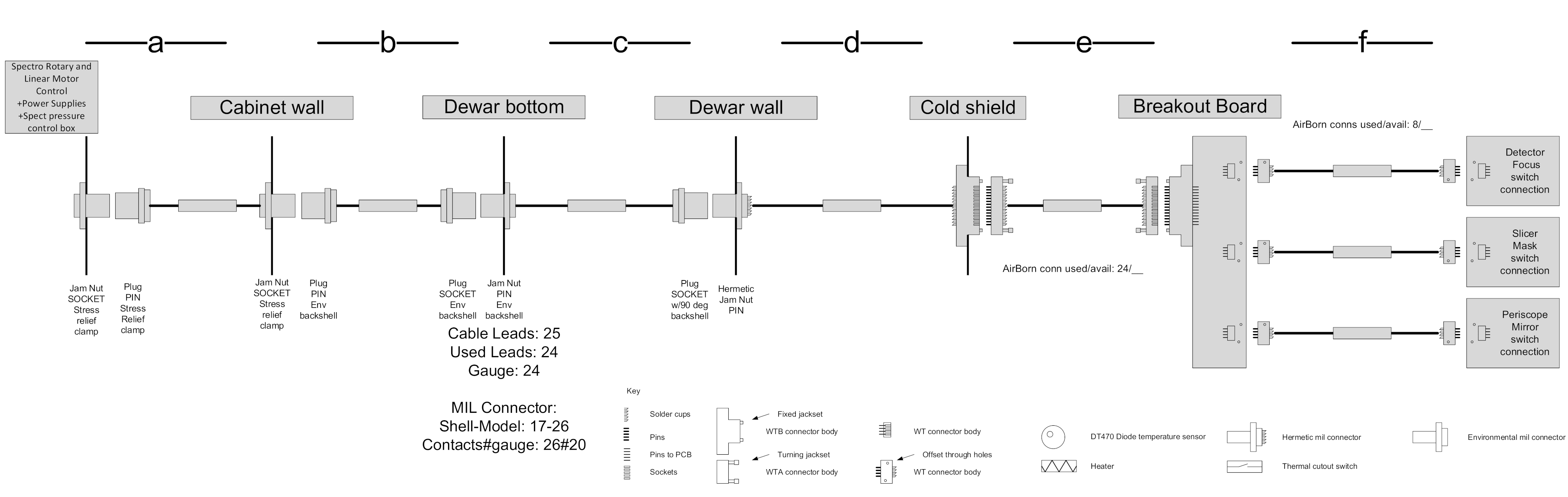}
    \end{tabular}
    \end{center}
    \caption[cable] 
    { \label{fig:cable} 
The average cable goes through 5-6 cable breaks on its way from the electronics cabinet, through the cable wrap, and into the science cryostat. Cables will be labeled for ease of organization and installation. For example, all cables between the cryostat wall and cold shield will have ``d" in their designation, cables going through the cable wrap will have ``b" in their designation, etc.}
    \end{figure}

IRIS is a large cryogenic cylinder 2 meters in diameter, 3.1 meters in height, with a total mounted mass of $\sim5400$ kg. It is suspended from underside of TMT's adaptive optics system NFIRAOS. IRIS will spin about its vertical axis to correct for field rotation. Most IRIS control electronics will be housed in a climate-controlled cabinet (called the nasmyth cabinet) on a platform below IRIS. Cables exit the nasmyth cabinet and go through a cable wrap, then up to a bulkhead attached to the underside of the cryostat. From there, the cables penetrate IRIS via multiple bulkheads on the sides of the cryostat (see Fig.~\ref{fig:cryostat}). Detector controllers are mounted on the sides of the cryostat in $\text{CO}_2$ gas cooled enclosures. 
In section \ref{sec:Arch}, we discuss some of the architectural challenges of IRIS with regards to the electronics system, section \ref{sec:hard} briefly discusses the nasmyth cabinets as well as connectors and cables we plan to use, and section \ref{sec:future} discusses some of the actions we plan to take in the near future.

\section{Architectural Challenges}
\label{sec:Arch}
A spinning, multi-ton object with over 50 cable attachments poses some design problems, many of which have been solved for smaller or simpler instruments. Integrating these solutions and coming up with new ones for such a large instrument is a challenge.

\subsection{Cryostat Rotation}
The instrument remains vertical for all observations, but must rotate about the vertical axis to correct for field rotation. Fig.~\ref{fig:cryostat} shows the limited space for the cable wrap and hoses. The cable wrap must be very complex and efficient for two reasons: (i) The radii of curvature of many cables in the wrap are near the radius of the cable wrap itself, and (ii) nearly all available space in the wrap will be filled due to the large number of cables. The wrap must be designed to impart minimal torque on the cryostat, as precision rotation is required, and minimal vibration can be conducted to the cryostat or AO system.

\subsection{Cryostat-Mounted Controllers}
Due to the long cable distance between the instrument cryostat and nasmyth cabinet, the Astronomical Research Cameras (ARC\cite{ARCinc}) controller system for the imager, spectrograph, and OIWFS must be placed directly on the cryostat or OIWFS in their own cooled enclosures. Mounting control electronics directly to the cryostat also reduces the amount of control cabling through the wrap, but at the same time, these boxes must be cooled via insulated $\text{CO}_2$ gas coolant lines that take up room in the wrap. Fig.~\ref{fig:leach} shows an image of the type of box to be mounted to the side of the science cryostat. The science detectors are connected directly to the control boxes via a $\sim 1$ meter ribbon cable. The boxes will contain amplifiers and analog-to-digital converters, as well as bias control, clocking, and other control electronics. This scheme is believed to slightly out-perform a cryogenic ASIC in terms of noise, and the complicated read-out patterns planned for the detector will be easier to implement with custom electronics.

\subsection{Cable Number/Length}
Fig.~\ref{fig:over} shows the large amount of cables ($\sim50$) between the nasmyth cabinet and the cryostat/OIWFS. On top of large numbers, the average cable goes through 5-6 cable breaks on its way (see Fig.~\ref{fig:cable}), bringing the number of individually fabricated cables to $\sim300$. The production and installation of this many long, heavy cables will be a challenge; the connections at cable breaks must be simple to assemble and disassemble on-site, and must be robust and repeatable. Simple steps like efficient naming schemes for cables will increase ease of organization and installation.

    \begin{figure} [hb]
    \begin{center}
    \begin{tabular}{c} 
    \includegraphics[width=15cm]{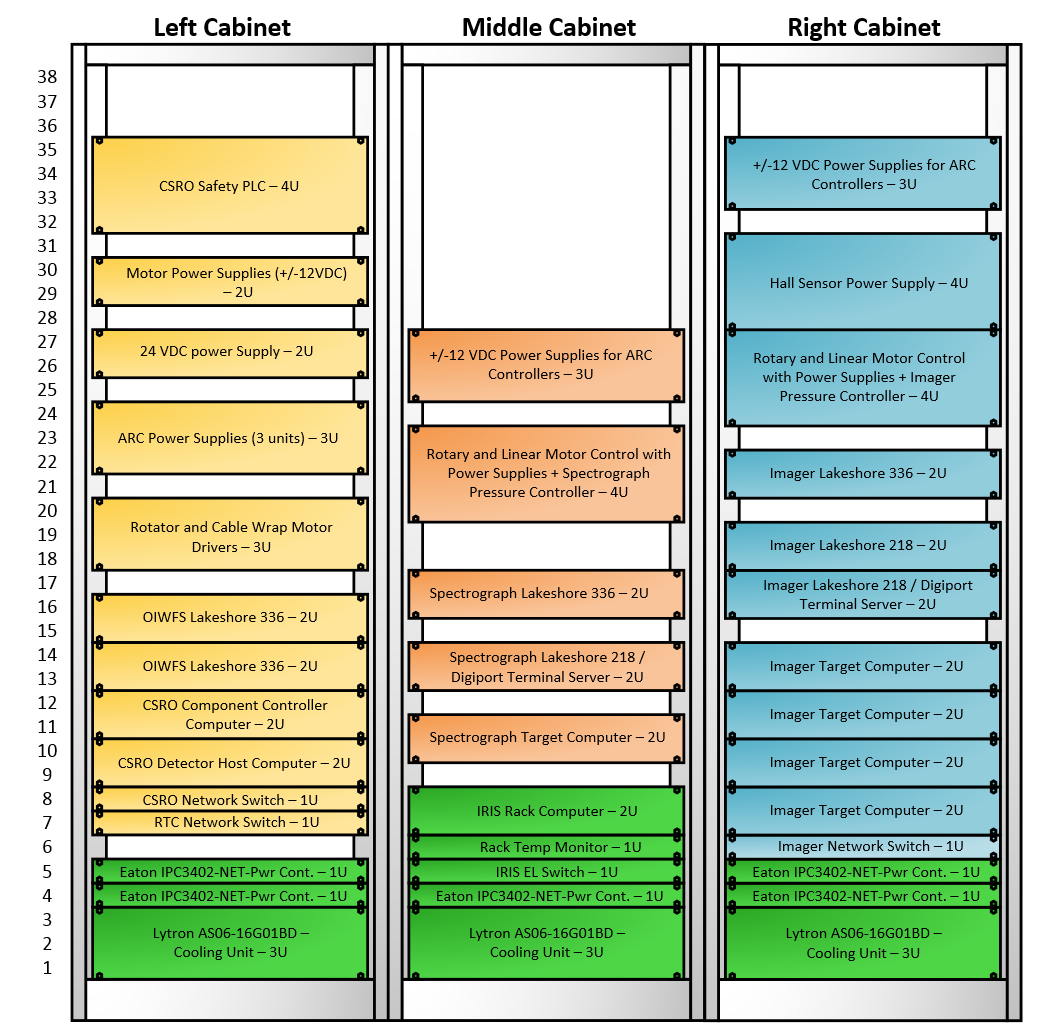}
    \end{tabular}
    \end{center}
    \caption[rack] 
    { \label{fig:rack} 
The layout for the electrical control components in the nasmyth cabinets located on the nasmyth platform (the same platform the cable wrap is located on; see Fig.~\ref{fig:cryostat}). There will be three cabinets, one devoted to each of the OIWFS, the spectrograph, and the imager. Sometimes a component will be racked by itself, like the ``Spectrograph Lakeshore 336" in the ``Middle Cabinet", and sometimes  components will be grouped in a custom box and racked together, like the ``Rotary and Linear...Pressure Controller" in the ``Middle Cabinet". Yellow items are OIWFS, cable wrap, and rotator components; orange are spectrograph components; blue are imager components; and green are cabinet environment control, power, and communication.}
    \end{figure}

    \begin{figure} [ht]
    \begin{center}
    \begin{tabular}{c} 
    \includegraphics[width=15cm]{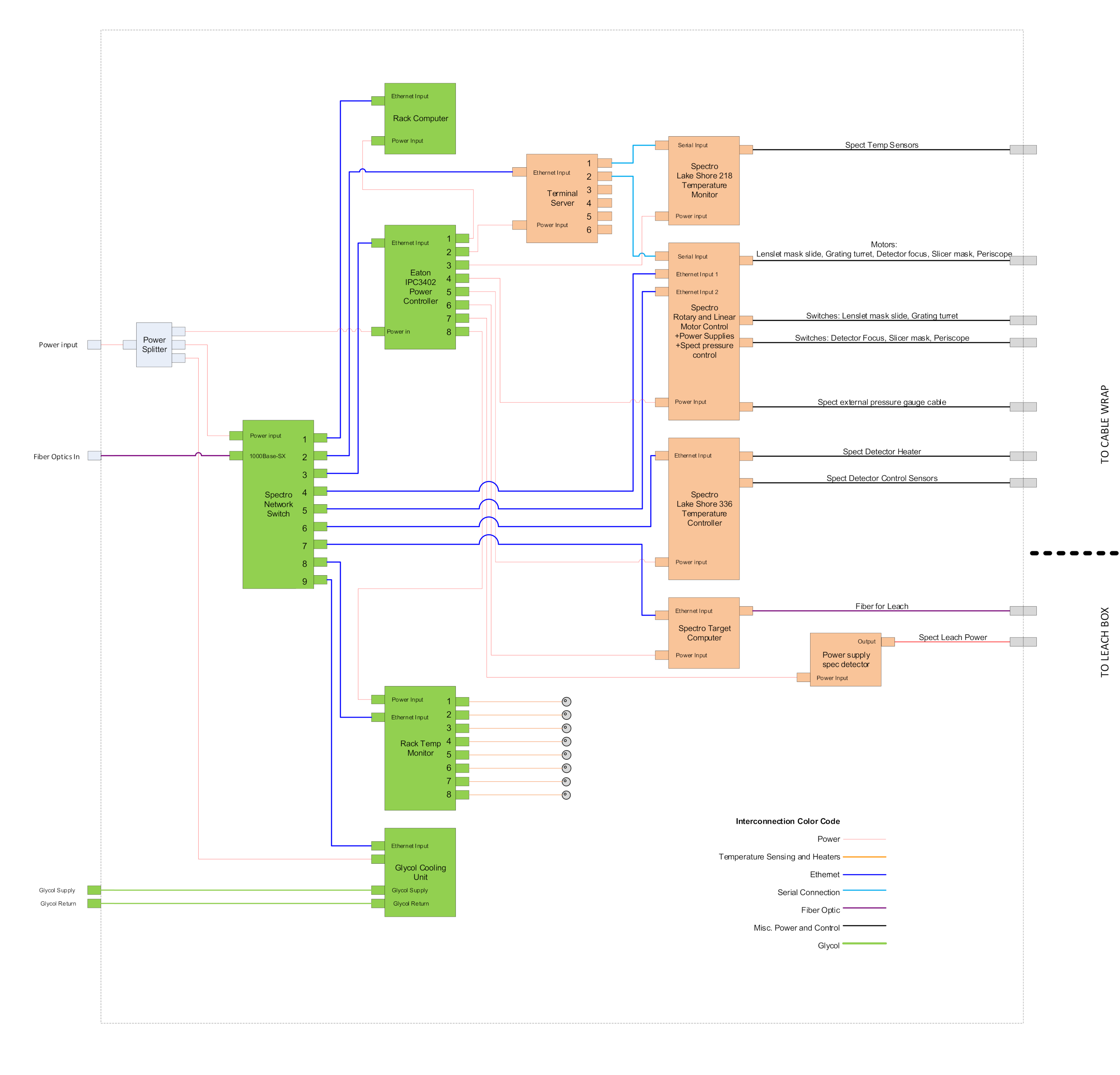}
    \end{tabular}
    \end{center}
    \caption[specab] 
    { \label{fig:specab} 
The architecture and type of connections between the various components of the ``Middle Cabinet" in Fig.~\ref{fig:rack}. Orange components are science cryostat control electronics, and green are cabinet environment control, power, and communication.}
    \end{figure} 
    
\section{Hardware}
\label{sec:hard}

\subsection{Nasmyth Cabinets}
\label{sec:nasm}

A large amount of control electronics are housed nearby the cryostat in three 19-inch rack cabinets, called nasmyth cabinets; each will control one of the OIWFS, Imager, and Spectrograph. Each cabinet has its own temperature monitor and glycol cooling unit for environment control. Fig.~\ref{fig:rack} shows the layout of various components in the nasmyth cabinets. Sometimes a component will be racked by itself, like the ``Spectrograph Lakeshore 336" in the ``Middle Cabinet", and sometimes  components will be grouped in a custom box and racked together, like the ``Rotary and Linear...Pressure Controller" in the ``Middle Cabinet". Fig~\ref{fig:specab} shows the architecture and type of connections between the various components of the ``Middle Cabinet" in Fig.~\ref{fig:rack}. 

\subsection{Cables and Connectors}
\label{sec:con}
Military-style connectors  are used in most non-cryogenic connections. They are durable enough to withstand the torques of the cable wrap, and can be keyed such that erroneous connections are impossible. AirBorn connectors are used for most cryogenic connections. They are simple to fabricate, and connect/disconnect easily; they also maintain good connections when cooled to cryogenic temperatures. 

Twisted/shielded pair cables will be used outside the cryostat. These cables provide shielding against EM interference as well as a high level of customizability in numbers of wires and wire gauges. Inside the cryostat, we will use constantan cables because of their performance at cryogenic temperatures.

\section{Future Work}
\label{sec:future}
We must determine the exact lengths, cable types, and connector specifications for each cable. The design of the $\text{CO}_2$ gas coolant system is still preliminary. Coolant pipes typically place more stringent constraints on cable architecture because of increased bend radius, diameter, and mass per meter.

\acknowledgments 
 
The TMT Project gratefully acknowledges the support of the TMT collaborating institutions. They are the
California Institute of Technology, the University of California, the National Astronomical Observatory of Japan,
the National Astronomical Observatories of China and their consortium partners, the Department of Science and
Technology of India and their supported institutes, and the National Research Council of Canada. This work
was supported as well by the Gordon and Betty Moore Foundation, the Canada Foundation for Innovation, the
Ontario Ministry of Research and Innovation, the Natural Sciences and Engineering Research Council of Canada,
the British Columbia Knowledge Development Fund, the Association of Canadian Universities for Research in
Astronomy (ACURA), the Association of Universities for Research in Astronomy (AURA), the U.S. National
Science Foundation, the National Institutes of Natural Sciences of Japan, and the Department of Atomic Energy
of India.



\end{document}